\documentclass[twocolumn]{aastex63}
\usepackage{graphicx}
\usepackage{amssymb}
\usepackage{amsmath}
\usepackage{hyperref}
\usepackage{lineno}

\def\gtrsim{\mathrel{\hbox{\rlap{\hbox{\lower4pt\hbox{$\sim$}}}\hbox{$>$}}}}
\def\lesssim{\mathrel{\hbox{\rlap{\hbox{\lower4pt\hbox{$\sim$}}}\hbox{$<$}}}}
\def\gtrsim{\mathrel{\hbox{\rlap{\hbox{\lower4pt\hbox{$\sim$}}}\hbox{$>$}}}}

\def\nustar{{\sl NuSTAR}}

\shorttitle{Hard X-ray spectrum of the Vela pulsar and its wind nebula constrained by NuSTAR}
\shortauthors{Kargaltsev et al.}

\begin{document}

\title{Hard X-ray spectrum of the Vela pulsar and its wind nebula constrained by NuSTAR}

\author{Oleg Kargaltsev}
\affiliation{Department of Physics, The George Washington University, 725 21st St. NW, Washington, DC 20052}

\author{Jeremy Hare}
\affil{Astrophysics Science Division, NASA Goddard Space Flight Center, 8800 Greenbelt Rd, Greenbelt, MD 20771, USA}
\affiliation{Center for Research and Exploration in Space Science and Technology, NASA/GSFC, Greenbelt, Maryland 20771, USA}
\affiliation{The Catholic University of America, 620 Michigan Ave., N.E. Washington, DC 20064, USA}

\author[0000-0003-3540-2870]{Alexander Lange}
\affiliation{Department of Physics, The George Washington University, 725 21st St. NW, Washington, DC 20052}
\affiliation{Department of Space Physics, University of Maryland Baltimore County, Baltimore, MD 21250, USA}

\begin{abstract}
We present the analysis of 200-ks NuSTAR observation of the Vela pulsar and the pulsar wind nebula (PWN). The phase-resolved spectra corresponding to two main peaks in the folded pulse profile differ significantly. The spectrum of Peak 1 is significantly harder than that of Peak 2 in qualitative agreement with the earlier RXTE results. However, for both spectra, the  values of power-law (PL) fit  photon  indices, $\Gamma$, are noticeably larger than the previously reported values. The  harder (Peak 1) spectrum has  $\Gamma=1.10\pm0.15$ which is close to those  
measured for the bright inner jets of the PWN.
 We used the off-pulse interval to  remove the  emission from the pulsar and measure the compact pulsar wind nebula (PWN) spectrum in hard X-rays.  
 We also measured the spectrum from the 
south-western (SW) region of the PWN which is resolved by NuSTAR from the compact PWN. For both  
regions, we fit the NuSTAR spectra by themselves and together with the Chandra X-ray Observatory spectra.
 We found that the  
  PWN spectrum (for both regions) requires a more complex model than a simple PL. The fits to compact PWN spectrum favor  exponentially cutoff PL model, with $E_c\approx 50$ keV over the broken PL model. The observed synchrotron photon energies imply electrons accelerated to $\approx$ 150 TeV.

\end{abstract}

\section{Introduction}
The Vela pulsar (PSR 0833-45), lying at a distance of  $\approx290$ pc \citep{2003MNRAS.343..116D}, is the youngest of the few nearby pulsars and has been studied extensively across the electromagnetic spectrum (see e.g., \citealt{2005ApJ...627..383R} and references therein).
 Spinning with a period of 89 ms, Vela is the brightest GeV  pulsar in the $\gamma$-ray sky.  The pulsar's spectrum is dominated by non-thermal emission everywhere but in soft X-rays where thermal emission from the neutron star (NS) surface prevails \citep{2001ApJ...552L.129P}. Outside the soft X-ray energy range \citep{2002ASPC..271..353S} the phase-integrated multiwavelength spectrum 
 is 
complex  
 showing multiple breaks (see e.g., \citealt{2011MNRAS.415..867D}). The multiwavelength pulse profiles are similarly complex with 
 5-6 discernible peaks throughout the 89-ms pulse period whose relative strengths vary with energy (see Figure 5 of \citealt{2019MNRAS.482..175S}).  These changes are noticeable even across the 2--30 keV  energy range explored with  the Rossi X-ray Timing Explorer (RXTE; \citealt{1999ApJ...524..373S,2002ApJ...576..376H,2015MNRAS.449.3827K}). By performing phase-resolved analysis of RXTE data   \cite{2002ApJ...576..376H} 
 found that the peak's spectra can be described by %an extremely hard
power-law model with photon indices, $\Gamma$ ranging from 0.8 to 2.1. The lowest value of  $\Gamma$ indicates a very hard spectrum 
 implying a very unusual particle spectral energy distribution (SED) assuming the standard 
synchrtoron emission mechanism. This spectrum is also harder than the spectrum of the hardest parts of the pulsar wind nebula (PWN), $\Gamma\approx 1.2$  \citep{2004IAUS..218..195K,2017JPhCS.932a2050K}.

Owing to its young age\footnote{ The age of 11 kyr is based on the standard assumption of a braking index of 3. The actual Vela pulsar age may be somewhat  larger (see, e.g., \citealt{2017MNRAS.466..147E}}.
 ($\approx 11$ kyrs) and relatively high spin-down power ($\dot{E}=7\times10^{36}$ erg s$^{-1}$), the pulsar inflates a bubble filled with  energetic particles within the Vela  supernova remnant (SNR) interior \citep{1998AJ....116.1886B,2003MNRAS.343..116D}. The compact axisymmetric PWN exhibits a complex fine structure (both spatial and spectral) resolved in the  Chandra X-ray Observatory (CXO) images \citep{2001ApJ...554L.189P,2001ApJ...556..380H,2003ApJ...591.1157P,2013ApJ...763...72D}.
 The PWN spectral maps, obtained from CXO data in the 0.5-8 keV range, 
 show one of the hardest X-ray spectra among all known PWNe \citep{2017JPhCS.932a2050K}.    Studying the spectral changes with distance from the pulsar constrains particle transport mechanisms and the evolution of the particle SED due to radiative cooling (see, e.g., \citealt{2016MNRAS.460.4135P}). Previous observations in hard X-rays lacked the angular resolution provided by NuSTAR \citep{2011ApJ...743L..18M}, making it difficult to isolate the PWN's contribution from the pulsar and to accurately subtract the local background.   The recent IXPE measurements found a very high degree  of polarization  (up to 60\%) for the compact PWN, suggesting  the electrons in
the PWN are accelerated with little or no turbulence
in a highly uniform toroidal magnetic field \citep{2022Natur.612..658X}.

 On larger  angular scales the PWN  is asymmetric, extending much further away from the pulsar in the SW direction compared to the north-eastern (NE) side. Unfortunately, IXPE does not provide polarization measurements outside the bright compact PWN but radio data still suggest that the magnetic field is highly regular and the polarization fraction remains high \citep{2003MNRAS.343..116D}. 
 The SW extension of the compact PWN connects smoothly to the so-called ``cocoon'' (a possible relic PWN), whose soft X-ray spectrum, measured with XMM-Newton, appears to be a mixture of thermal and non-thermal emission \citep{2018ApJ...865...86S}. The cocoon is also detected at TeV energies with the High Energy Stereoscopic System (H.E.S.S.;   \citealt{2019A&A...627A.100H}).

We obtained NuSTAR observation of the Vela PWN to extend the spectral coverage out to 70 keV for the compact PWN, to separate the pulsed emission from that of the PWN using photon phases, and to spatially resolve the spectrum of the SE part of the PWN from that of the  compact PWN.   We also made use of Fermi Large Area Telescope (LAT) data to determine pulsar's spin period and its spin period derivative and to compare $\gamma$-ray and hard X-ray pulse profiles.  In Section 2 we describe the data used in this paper while Sections 3 and 4 report the results of timing and spectral analysis, respectively.  The discussion of the results follows  in Section 5. We conclude with the summary in Section 6. 

\section{Observations}

\subsection{NuSTAR}

\nustar\ \citep{2013ApJ...770..103H} observed the Vela pulsar region on 2021 March 21 (ObsID  30601032002; start time MJD %59294.66333575769 
59294.66253502, stop time MJD 59298.99801064)  with a total scientific exposure of 205,080 seconds. 
We reprocessed the data using {\tt HEASoft} v6.29 and \nustar\ CALDB v20210427.  
We ran the standard  {\tt nupipeline} tool, which applied all the latest calibrations and filtering. We also barycenter-corrected the photon arrival times originating from the pulsar's position, using \nustar\ clock correction file 20100101v125 which also corrects for \nustar\'s clock drift, providing a timing accuracy of $\sim$65 $\mu$s \citep{2021ApJ...908..184B}.  
We excluded times of high particle background{\bf\footnote{See \url{https://nustarsoc.caltech.edu/NuSTAR_Public/NuSTAROperationSite/SAA_Filtering/nulyses_reports/30601032002/nu30601032002_SAA_Report_A.pdf}}} using the options {\tt saacalc=2}, {\tt saamode=optimized}, and {\tt tentacle=no}, which had a minimal impact on the total exposure time 
($<0.2\%$).
We extracted spectra using {\tt nuproducts} (with option {\tt extended=yes} for analysis of the  PWN).

\subsection{Chandra X-ray Observatory}

We also used archival CXO ACIS imaging data (from 3 observations obtained between Aug 15 2010 and Sep 4 2010; ObsIDs 12073, 12074, and 12075; total scientific exposure 121 ks) to improve the statistics and to extend the data analysis to lower energies  for the PWN spectra. The data reduction and fits followed standard procedures described in {\bf the} CIAO  analysis threads\footnote{\url{https://cxc.cfa.harvard.edu/ciao/threads/}}. Since all three CXO observations are taken close to each other and the target was placed at the same location on ACIS-S3 chip, we chose to combine\footnote{Following the recipe provided at  \url{https://cxc.cfa.harvard.edu/ciao/threads/coadding/}} spectra and responses for all observations
 to simplify the subsequent analysis steps. However, we also verified that including all 3 CXO spectra individually produces similar results. For our analysis, we used CIAO  4.13 and CALDB 4.9.4. 

\subsection{Fermi LAT}

More than 13 years of Fermi LAT data were collected and used to analyze the high-energy emission from the Vela pulsar. We used \emph{gtselect} from \emph{FermiTools}{\bf\footnote{\url{https://fermi.gsfc.nasa.gov/ssc/data/analysis/software/}}} to extract data from a 1\textdegree \, radius region of interest centered on Vela pulsar. Additionally, we applied a selection on the times and energies, filtering our dataset to match the Fermi LAT ephemeris range (valid between MJD  54682.71577-59500.98129) and limiting photon energies to be between 0.3 GeV to 6 GeV. Barycenteric corrections were made with PINT's software \citep{2021ApJ...911...45L}. We used the  JPL  DE421  planetary ephemeris 
for all timing analyses.

\section{Timing}
\label{timing}

\begin{figure}
\centering
\includegraphics[trim={0 0 0 0},scale=0.3]{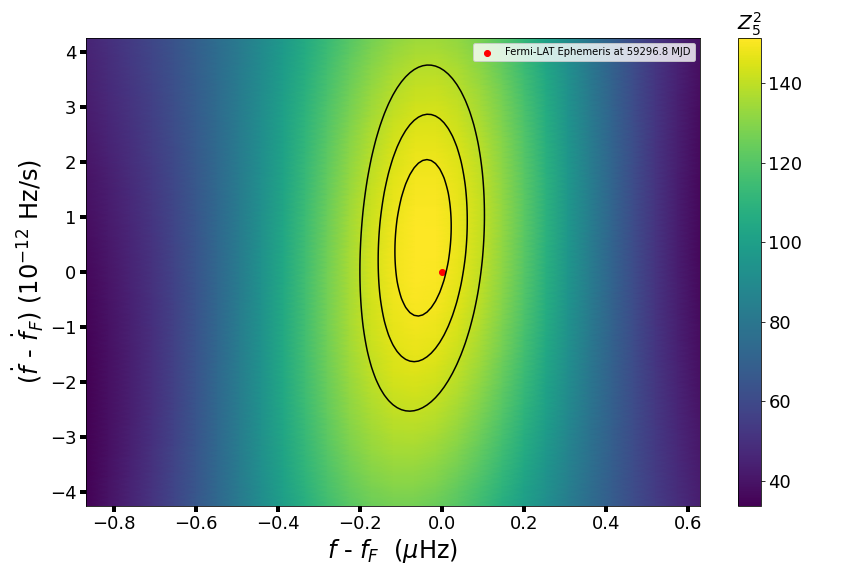}
\caption{$Z_5^2$ distribution as a function of frequency and its derivative for \nustar\ data. Zero (shown by a red point) corresponds to $f=f_F=11.18437086731769$ Hz and $\dot{f}=\dot{f}_F=-1.52431976409 \times 10^{-11}$  Hz s$^{-1}$ inferred from the Fermi LAT ephemeris for the middle of \nustar\ observation (MJD 59296.83198754).  The black lines show 1, 2 and $3\sigma$ confidence contours  (see   \citealt{1983A&A...128..245B} and Appendix A in \citealt{2024arXiv240704337P}. }
\label{fig:zsq}
\end{figure}

We performed $Z_n^2$ test (\citealt{1983A&A...128..245B}; for n=5, given the multi-peak nature of the hard X-ray pulse profile; \citealt{2002ApJ...576..376H})  on the NuSTAR data, using a 2D grid of frequency ($f$) and frequency derivative ($\dot{f}$). We also varied the circular  extraction  aperture radius,  to maximize the pulsar contribution, and found that the $Z_5^2$ values grow up to $r=50''$ 
 and then level off. In addition, we explored the dependence of  $Z_5^2$ on the energy range and found that it is maximized  when the full range (3-79 keV) is used.
Figure \ref{fig:zsq}   shows the expected    
$f$ and $\dot{f}$ 
 calculated using the   \emph{PINT} timing software  
with the Vela Fermi LAT ephemeris\footnote{The ephemeris  validity range  MJD  54682.71577-59500.98129 includes the \nustar\ observation interval.} based on 13.2 years of data (courtesy of Mathew Kerr). 
 Figure \ref{fig:zsq} shows that the prediction of $f$ and $\dot{f}$ from \nustar\ timing agrees with that based on the Fermi LAT ephemeris.

To produce the \nustar\ pulse profiles we adopted two different approaches. In the first approach, we assigned phases to all \nustar\ photons extracted from the $r=50''$ aperture (in 3-79 keV band),  
as well as the Fermi-LAT photons (in 0.3-6 GeV band), using PINT's \emph{photonphase} and \emph{fermiphase} respectively.  
This approach ensures that the \nustar\ and LAT profiles are co-aligned (Figure \ref{fig:nustar_lat_profiles}) and shows that the two main peaks, dubbed Peak 1 (Pk1) and Peak 2 (Pk2), seen in the \nustar\ data coincide with the peaks seen in the Fermi LAT pulse profiles. 
In the other approach, we used the $f$ and $\dot{f}$ values   obtained\footnote{for the middle of the \nustar\ observation (59296.84506887 MJD).} from the $Z_5^2$ test to fold the \nustar\ pulse profile. As one can see from Figure \ref{fig:nustar_lat_profiles} there are small differences between the two NuSTAR profiles (first and second panels from the top) but overall they are quite similar.  The corresponding pulsed fractions (defined as the ratio of the area above the minimum to the total area under the pulse profile) are $4.0\%\pm1.1\%$ and $4.5\%\pm1.1\%$ , respectively.  In the same figure we also plot pulse profiles in 3-10 and 10-79 keV. They are fairly similar with Pk2 becoming broader and less pronounced in 10-79 keV, in agreement with its softer spectrum (see Section \ref{Timing_Pulsar}).

\section{Spectra}

 In this section we describe spectral fitting  performed with XSPEC package version 12.13.1 \citep{1996ASPC..101...17A}. 
 
\subsection{Pulsar}
\label{Timing_Pulsar}

Since the PWN photons strongly dominate those from the pulsar  within the chosen $50''$ aperture centered on the pulsar, we only fit the pulsar spectrum extracted for the two phase bins corresponding to  
Pk1 and 
Pk2 in the $\gamma$-ray pulse profile (see Figure \ref{fig:nustar_pulse} and Table \ref{tab:phs-resolved}). 
 Since the intervening hydrogen column is small ($\sim3\times10^{20}$ cm$^{-2}$), and NuSTAR observes at energies $>$ 3 keV, we simply  ignore the foreground absorption.
 The PWN emission strongly dominates the pulsed emission, so to account for it when fitting the spectra of the peaks, we first fit the spectra of the Off-Pulse (OP) interval ($\phi=0.74 - 0.80$) with a power-law. We then fit the  Pk1 ($\phi=0.18 - 0.22$) and Pk2 ($\phi=0.52 - 0.68$) spectra with two PLs, where the for the first PL the photon index and normalization are frozen to their best-fit values from the fit the to OP spectra while for the second PL (representing the Pk1 or Pk2  spectra) both parameters are fitted. We find the Pk1 spectrum with $\Gamma_1=1.07\pm0.15$ to be  substantially harder than the Pk2 spectrum with $\Gamma_2=1.62\pm$ 0.20. The uncertainties are given at the 68\% confidence level for a single interesting parameter.
 The PL fit quality is good with $\chi^2/{\rm d.o.f.}\approx0.90$ and $1.16$) and for  Pk1 and Pk2 spectra, respectively.

 \begin{figure}
\centering
\includegraphics[trim={0 0 0 0},height=20.7cm]{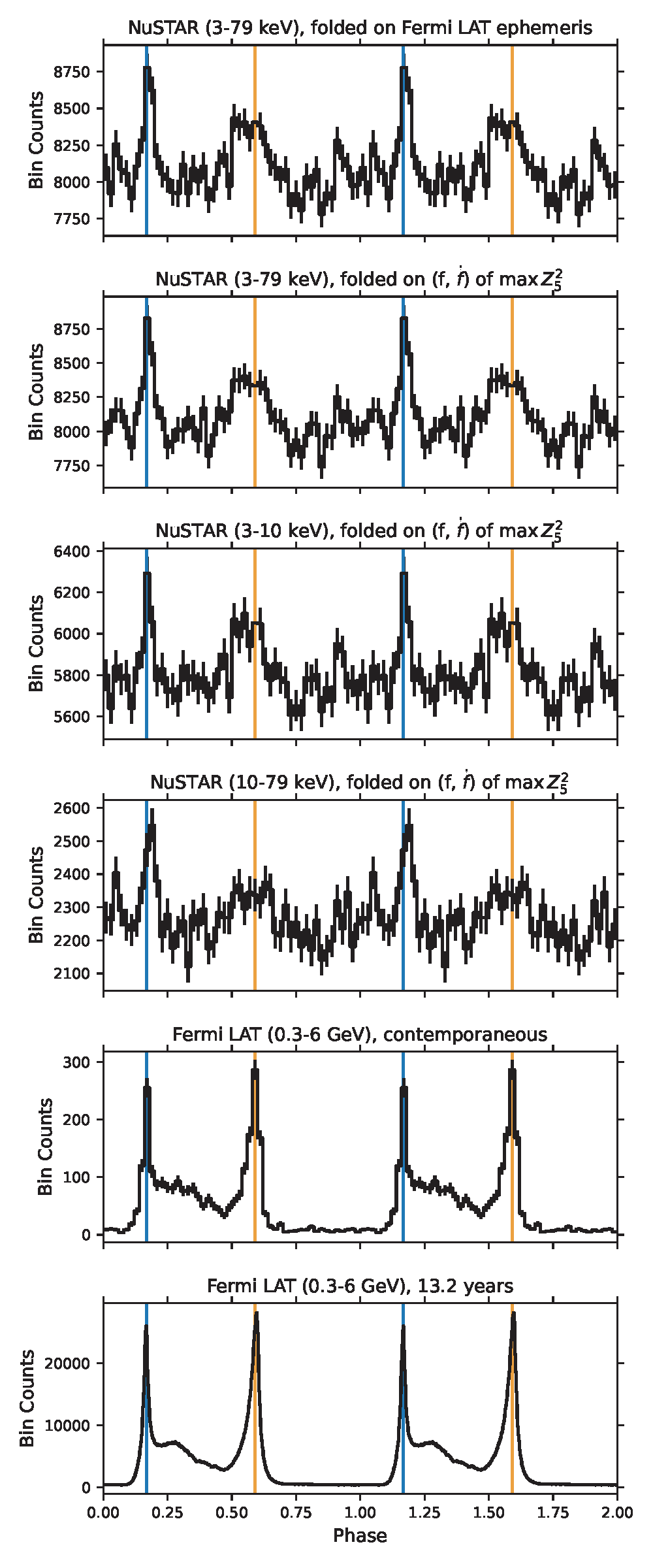}
\caption{
NuSTAR and Fermi LAT pulse profiles. 
 Each pulse NuSTAR profile consists of 50 equal-width phase bins.  
 The blue and orange vertical lines correspond to the phase of  Peak 1 (Pk1) and Peak 2 (Pk2) as determined from the plot in the bottom panel. See Section \ref{timing} for details. The epoch of zero phase is 56623.15526604 MJD.
\label{fig:nustar_lat_profiles}
}
\end{figure}

\subsection{PWN: NuSTAR}

In order to explore the PWN spectrum we define two regions, one for the compact PWN and one for the SE extension (white and green circles in  Figure \ref{fig:images}, respectively).  The results of the fits are summarized  in Table \ref{tab:pwn_fits}. The extracted spectra are binned to $\ge$200 photons per bin.    In both cases the background spectrum is extracted from the  $r=40''$ circle  located west  of the pulsar, on the same chip as PWN,  centered at R.A.=8h34m54s, Decl.=-45$^{\circ}12'04''$ (see Figure \ref{fig:images}). During the spectral fits the background spectrum is subtracted from that of the source after being binned the same way. 
We first fit \nustar-only spectra (in 3-79 keV range). The photon indices are tied for FPMA and FPMB detector spectra but their normalizations are left free.

  For the compact PWN region ($r=50''$ circle centered on the pulsar) we only use 
off-pulse (OP)  photons  to avoid contamination by the pulsar  (99\% of photons are expected to come from the PWN). 
The spectrum 
contains 9,661 (FPMA) and 10,557 (FPMB)
 photons in 3-79 keV. The background  contribution  is negligible. 
 The best-fit {with a simple PL model} gives $\Gamma=1.64\pm0.01$ but the fit quality is poor with $\chi^2=$141  for 91 d.o.f. Also, some systematic large-scale residuals at low and high energies of the  energy range are present. We, therefore, tried to also fit a broken PL model (``bknpower'' in XSPEC;  hereafter BPL) and obtained $\Gamma_1=1.57\pm0.02$, break energy $E_b= 10.9\pm1.5$  keV  and $\Gamma_2=1.79\pm0.05$ with a noticeable improvement in large-scale residuals resulting in $\chi^2=$128  for 89 d.o.f. An exponentially cutoff PL   (``cutoffpl'' in XSPEC;  hereafter ECPL) model fits   equally well ($\chi^2=$122  for 91 d.o.f.) with $\Gamma=1.50\pm0.04$ and a rather uncertain exponential cutoff energy $E_c=79^{+26}_{-16}$ keV.

The spectrum from SW extension region (see Figure \ref{fig:images}, bottom)
has 17,059 and 16,380 photons in FPMA and FPMB, respectively,  with 80\% coming from the source for either of the detectors.   The PL fit quality is good with $\Gamma=1.81\pm0.02$  with $\chi^2=232$ for 241 d.o.f.
There are no significant systematic residuals. 
Therefore, we have not attempted any fits with more complex models.

\begin{figure}
\centering
\includegraphics[trim={0 0 0 0},scale=0.6]{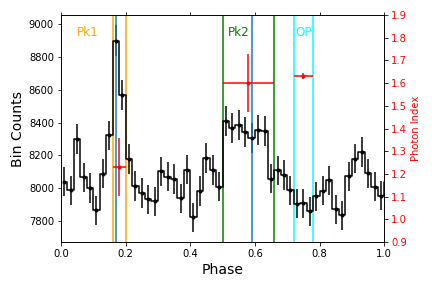}
\caption{The black line and errorbars show the 3-79 keV NuSTAR pulse profile of Vela pulsar with the 3 phase intervals used for the phase-resolved analysis shown (orange - Pk1; green - Pk2, blue - OP). The dark blue vertical lines show the phases of the two peaks in the Fermi LAT pulse profile.  The red errorbars show the photon index measurements (on the  right vertical axis) for the corresponding phase intervals. See also Table \ref{tab:phs-resolved} and text for details.
\label{fig:nustar_pulse}
}
\end{figure}

\subsection{PWN: NuSTAR+CXO}

We now perform the joint fits to the \nustar\ and CXO ACIS spectra for the same as above two PWN regions.  For CXO observations the background is taken from a source-free circular region with $r=1'$ located north of the pulsar and centered at R.A.=8h35m18s and Decl.=--45$^{\circ}07'10''$.  We   jointly fit spectra  in 1-79 keV range with the CXO ACIS spectrum extending up to 5 keV and NuSTAR spectrum starting from 3  keV. We also excluded a narrow 1.9-2.3 keV region 
%around 2 keV
%in
from the CXO spectrum  because it seems to show local residuals which could be instrumental\footnote{ The residuals are stronger in two out of 3 CXO observations. They become noticeable only with a very large number of photons in the fit to the compact PWN region and are only marginally seen in the fit for the SW extension. However, for consistency, we excluded the same energy range in both spectra. }. All spectra are binned to have at least 200 counts per bin. For both CXO and NuSTAR the background spectrum is subtracted from the source's spectrum.
In these fits we used models modified by ISM absorption modeled with XSPEC's ``tbabs''   using {\tt wilm} abundances \citep{2000ApJ...542..914W} with fixed $n_H=3.2\times10^{20}$ cm$^{-2}$ (see \citealt{2001ApJ...552L.129P}). 
 The normalizations of different detectors/instruments (FPMA, FPMB, and ACIS) are allowed to vary independently while the rest of the fitting parameters are tied.

\begin{figure*}
\centering
\includegraphics[trim={0 0 0 0},scale=0.48]{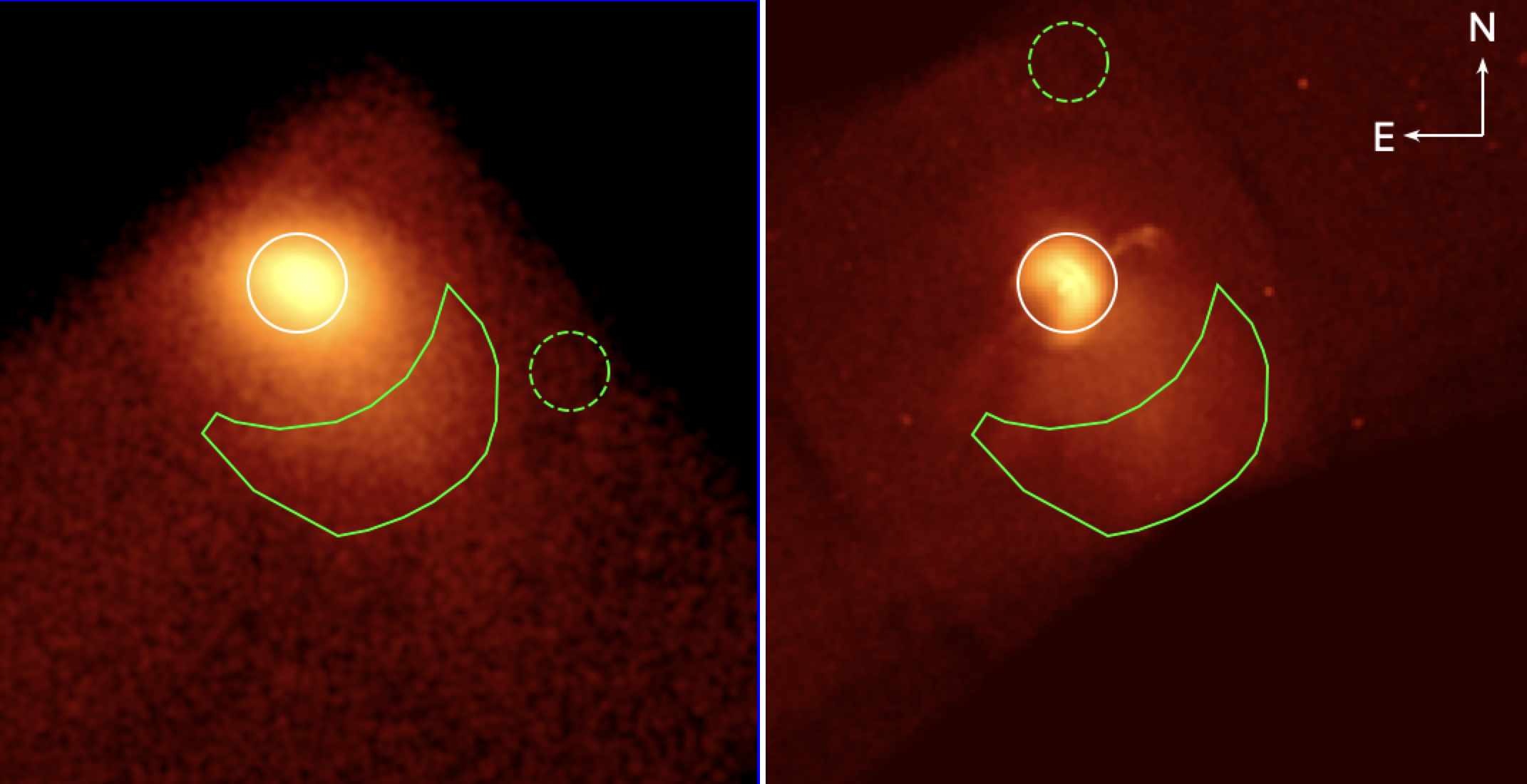}
\caption{{\sl NuSTAR} 3-79 keV (left) and {\sl CXO} {\bf 1-8 keV} (right) images of the Vela PWN showing the same region of  the sky.  Both images are binned to $2''$  pixel size and smoothed with $r=4''$ constant Gaussian kernel. The white ($r=50''$) and green ($r=1.2'$) circles show spectral extraction regions for the compact PWN  (with  exclusion of $r=3''$ pulsar region; not shown) and SW extension region, respectively. North is up, East is to the left.
\label{fig:images}
}
\end{figure*}

\begin{figure*}
\centering
\includegraphics[trim={0 0 0 0},scale=0.60]{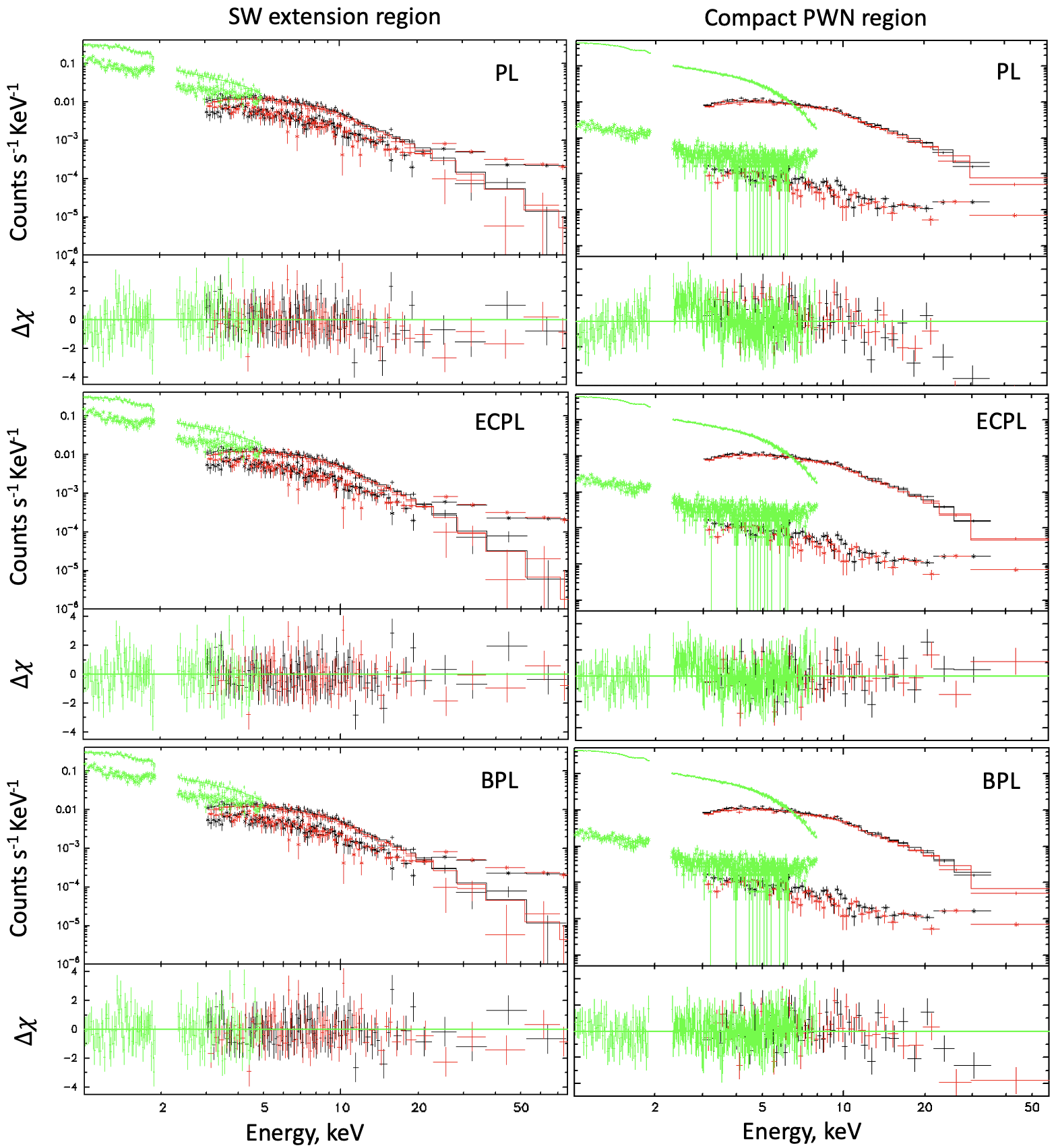}
\caption{Joint fits to NuSTAR (black and red; for FPMA and FPMB, respectively) and CXO (green) spectra. Background contribution is also shown for each detector/instrument.
\label{fig:nustar_cxo_spectra}
}
\end{figure*}

For the SW extension PWN region  (polygon if Figure \ref{fig:images}), the joint spectra and the best fits are shown in Figure   \ref{fig:nustar_cxo_spectra}, left panels. To avoid contamination by the compact PWN, the polygon boundary is at least $150''$ away from the pulsar and $\approx100''$ away from the bright compact PWN structure (arcs and jets) (seen in the CXO image). 
% jointly fit spectra  in 1-79 keV range\footnote{Below 1 keV there are relatively  few photons from the PWN compared to the instrument background because of the contamination accumulated on the ACIS optical light blocking filter.}. 
Although the single PL fit quality is acceptable, with $\Gamma=1.81\pm0.02$, there is a  noticeable large-scale structure in the residuals. Both BPL and ECPL models are providing good fits. For the former, $\Gamma_1=1.64^{+0.03}_{-0.05}$, break energy $E_b= 3.0^{+0.4}_{-0.7}$  keV and $\Gamma_2=1.82\pm0.02$. The low energy of the break could explain why the BPL model was not required when only \nustar\ data were fitted in the 3-79 keV range. On the other hand, the coincidence between the  $E_b$ and the start of \nustar\ energy range %may
indicate that the improvement associated with the broken PL may come from imperfection in calibration 
(see also \citealt{2017AJ....153....2M}).  
If the latter is indeed the case then one can expect a similar situation with spectral fits for the compact PWN region.  The ECPL model fits the spectrum nearly equally well with $\Gamma=1.64\pm 0.02$ and $E_c=50^{+14}_{-9}$ keV.

 For 
%this 
the compact PWN region region the CXO and NuSTAR spectra and the joint fits are shown in the right panels of Figure \ref{fig:nustar_cxo_spectra}.
 For the ACIS spectrum, also extracted from $r=50''$ region centered on the pulsar,  
 we exclude the $r=3''$ region centered on the pulsar.  
 We  use the OP phase interval for \nustar\ data in order to 
exclude the pulsar. 
 The simple absorbed PL fit is clearly not satisfactory (see Figure \ref{fig:nustar_cxo_spectra}). While the  BPL fit is 
better ($\chi^2$=526 for  458 d.o.f.), it does show systematic residuals at high energies with the model over predicting the data. The best-fit parameters are $\Gamma_1=1.46\pm0.01$, break energy $E_b= 2.7\pm0.1$  keV and $\Gamma_2=1.60\pm0.01$. The break energy is again close to the  lower energy end of \nustar\ range which may hint at a possible cross-calibration issue. The above-mentioned high-energy residuals disappear when the  broken PL model is replaced with the ECPL model ($\chi^2=537$ for 459 d.o.f.) with best-fit $\Gamma=1.459\pm0.005$ and the  cutoff energy of $E_c=50\pm4$ keV.  However,  the positive residuals between 2 and 3 keV become more noticeable. This could again be  an indication of imperfect cross-calibration which is more noticeable for the bright compact PWN due to the much larger number of photons in the ACIS spectrum.

\begin{deluxetable}{lcc}
\tablecolumns{5}
\tablecaption{Phase-resolved spectral fitting results.
} 
\label{tab:phs-resolved}
\tablewidth{0pt}
\tablehead{
\colhead{Feature name} & \colhead{Phase range} & \colhead{$\Gamma$}  
%\colhead{ arcsec} & \colhead{counts} & \colhead{Hz} & 10$^{-11}$\colhead{Hz/s}& \colhead{}% & \colhead{} 
}
\startdata
Narrow peak small (Pk1) & 0.16-0.20 & $1.06\pm0.16$\\
%Narrow peak large & 0.70-0.78 & --\\
Broad peak (Pk2) & 0.50-0.66 & $1.62\pm0.20$\\
Background (OP interval) & 0.72-0.78 &   1.65$\pm0.01$\\
\enddata
\end{deluxetable}

\begin{deluxetable*}{llllllllll}

\tablehead{ 
\colhead{Region} & \colhead{Model} & \colhead{Mission(s)}   & \colhead{$\Gamma$ or $\Gamma_1$} & \colhead{$E_c$ or $E_b$ (keV)} & \colhead{$\Gamma_2$} & \colhead{$\chi^2$} & \colhead{d.o.f.} 
}
\startdata
Comp.\ PWN  & PL & NuSTAR & $1.64\pm0.01$ & $-$ & $-$ & 141 & 91 
\\
Comp.\ PWN  & ECPL & NuSTAR & $1.50\pm0.04$ & $79^{+26}_{-16}$ & $-$ & 128 & 90 
\\
Comp.\ PWN & BPL & NuSTAR & $1.57\pm0.02$ & $10.9\pm0.5$ & $1.79\pm0.05$ & 122 & 89
\\
 Comp.\ PWN & PL & NuSTAR+CXO & $1.509\pm0.002$ & $-$ & $-$ & 733 & 460 
% Comp.\ PWN & ECPL & NuSTAR+CXO & $1.48\pm0.01$ & $58^{+7}_{-5}$ & ... & 543 & 473 
\\
Comp.\ PWN & ECPL & NuSTAR+CXO & $1.459\pm0.005$ & $50\pm 4$ & $-$ & 537 & 459 
\\
% Comp.\ PWN & BPL & NuSTAR+CXO & $1.45\pm0.05$ & $2.7\pm0.1$ & $1.60\pm0.01$ & 543 & 472 
Comp.\ PWN & BPL & NuSTAR+CXO & $1.46\pm0.01$ & $2.7\pm0.1$ & $1.60\pm0.01$ & 526 & 458 
\\
\hline 
SW PWN & PL & NuSTAR & $1.81\pm0.02$ & $-$ & $-$ & 232 & 241
\\
SW PWN & PL & NuSTAR+CXO & $1.73\pm0.01$ & $-$ & $-$ & 413 & 370 
\\
SW PWN & ECPL & NuSTAR+CXO & $1.64\pm0.02$ & $50^{+14}_{-9}$ & $-$ & 391 & 369 
\\
% SW PWN & BPL & NuSTAR+CXO & $1.61\pm0.02$ & $1.81\pm0.01$ & $3.0\pm0.3$ & ... & ... 
SW PWN & BPL & NuSTAR+CXO & $1.64^{+0.03}_{-0.05}$ & $3.0_{-0.7}^{+0.4}$ & $1.82\pm0.02$ & 389 & 368 
%\\
\enddata

\label{tab:pwn_fits}
\caption{ Spectral fits with 3 different models  for compact and SW PWN regions. See text for details. The uncertainties are given at the 68\% confidence level for a single interesting parameter.  For the CXO+NuSTAR fits we fixed $n_H$ at $3.2\times10^{20}$ cm$^{-2}$.}

\end{deluxetable*}

\section{Discussion}

\subsection{Pulsar}

The NuSTAR pulse profile of Vela pulsar is remarkably complex in 3-79 keV range  with multiple narrow peaks of which two  align   
with the two main peaks (Pk1 and Pk2) seen in the LAT pulse profile (see Figure \ref{fig:nustar_lat_profiles}). However, while in the LAT pulse profile Pk1 and Pk2 are both very sharp, the hard X-ray counterpart of Pk2 is much broader than that of Pk1. The  3-79 keV NuSTAR pulse profile  overall looks 
%very
similar to the one obtained from RXTE  PCA (Proportional Counter Array) data    in 8-24 keV range more than two decades ago. In particular, the four most pronounced peaks (two main peaks Pk1 and Pk2, and 2 smaller ones preceding Pk1) are easily identifiable in both  NuSTAR and RXTE pulse profiles (see Figure 4 of \citealt{2015MNRAS.449.3827K}).   We note that, while we find  a modest evolution in the pulse shape  between the 3-10  and 10-79 keV energy ranges (Figure \ref{fig:nustar_lat_profiles}), the RXTE pulse shape appears to evolve more substantially between 4-8  and 8-24 keV (e.g., relative amplitudes of Pk1 and Pk2 change noticeably; see Figure 4 in \citealt{2015MNRAS.449.3827K}). While we are not exactly sure about the reason of the apparent discrepancies, they may originate from different dependencies of the RXTE PCA and NuSTAR effective area dependencies on energy and the very different spectra of Pk1 and Pk2.

The most intriguing feature of the multiwavelength energy-resolved pulse profiles (shown in, e.g.,  Figure 4 of \citealt{2015MNRAS.449.3827K}) is the very different behavior of the  Pk1 and Pk2  amplitudes and shapes as a function of energy. It can be summarized concisely as follows.  Pk1  remains narrow and prominent across a very wide energy range from $\sim 4 $ keV to $\sim 1$ GeV but decays above this energy (see \citealt{2015MNRAS.449.3827K} and  \citealt{2023NatAs...7.1341H}). It also disappears (or shifts to a noticeably different phase) in UV and optical  (see, e.g., \citealt{2019MNRAS.482..175S}). On the other hand, Pk2 persists from  the optical to the highest gamma-ray energies   ($\sim100$ GeV; \citealt{2018A&A...620A..66H}; Lange, et al.\ in prep.) but becomes weak and broad in hard X-rays. Naively, one may think that the nature of emission (emission mechanism) responsible for  Pk2 changes from optical to gamma-rays with the 
%change 
 transition to IC 
happening somewhere in the hard X-ray band. On the other hand, the stability of Pk1  may indicate a single emission mechanism operating in a very broad range of energies. If both peaks are due to emission associated with the  particles accelerated in the current sheet outside the light cylinder (as suggested by recent models, e.g.,  \citealt{2015ApJ...801L..19P,2016MNRAS.457.2401C}) then the synchrotron emission from particles with a  very broad SED could be the predominant Pk1 emission mechanism.  In this case, Pk1 could be associated with positrons propagating outwards in the  current sheet  located farther away from the pulsar than the Y-point\footnote{  See, e.g., \citealt{2017SSRv..207..111C}, for a modern description of pulsar magnetospheres} while Pk2 would be the  (synchro-)curvature emission from electrons precipitating towards the star which could be more important at lower energies. The difference in  the evolution of Pk1 and Pk2 as a function of energy might also be associated with differences in the  direction of motion of the radiating particles (toward vs.\ away from NS)
 attributed to the different acceleration  history and different emission sites (Y-point vs.\ current sheet) as proposed by \cite{2016MNRAS.457.2401C}.  This may also explain the observed  significant difference between the slopes of NuSTAR spectra for Pk1 and Pk2. However, a more conclusive answer requires fitting the physical model  (such as the one proposed by \citealt{2016MNRAS.457.2401C}) to multiwavelength phase-resolved spectra (which is beyond the scope of this paper).

Due to the coarse NuSTAR resolution, the pulsar is not resolved well from the surrounding bright compact PWN. Therefore, the pulsed component makes only a small contribution on top of PWN emission in the NuSTAR pulse profile. This only allows us to extract meaningful spectra for the 2 main peaks  while using photons from the OP interval as the background. 
 Although the Pk1 spectrum ($\Gamma_1=1.10\pm0.15$) is substantially harder than the  Pk2 spectrum ($\Gamma_2=1.62\pm0.20$), in qualitative  agreement with  the \cite{1999ApJ...524..373S} and \cite{2002ApJ...576..376H} results, the actual best-fit values are different 
and significantly larger than those reported by  \cite{1999ApJ...524..373S},  ($\Gamma_{1, \rm RXTE}=0.68\pm0.14$ and $\Gamma_{2,\rm RXTE}=1.17\pm0.12$).   While the values reported in Table 2 of  \cite{2002ApJ...576..376H} are  closer to ours, their Pk1 photon index is still noticeably smaller.   Since RXTE PCA  is a non-imaging instrument 
the differences could be associated with  the 
background treatments in the RXTE data analysis. 
We note that the new, larger values, are comparable to the photon index values for the hardest elements of the compact bright PWN (see \citealt{2017JPhCS.932a2050K}).

\subsection{PWN}

Both by fitting just the NuSTAR spectrum and by jointly fitting the NuSTAR and CXO ACIS spectra we find that a single PL cannot adequately describe the spectrum of the compact PWN (extracted from an $r=50''$ circle centered on the pulsar). At least one spectral break is required\footnote{ While the detailed modeling PWN particle SED evolution is beyond the scope of this paper we note that one should expect the PL to become  steeper gradually as a function of distance from the pulsar (an example of calculation in cylindrical geometry is shown  in Figure 10 of \citealt{2006ApJ...651..237C}) if the spectra are extracted from the sufficiently  small  regions. However, due to the wide PSF or NuSTAR we  use large regions and then it is reasonable to expect the  gradually softening spectrum can be approximated by a broken PL.} to describe the softening of the spectrum with energy. However, the ECPL model provides an even better description of the spectrum at higher energies suggesting a cutoff energy of $\sim$50 keV. This  
indicates that the radiative cooling 
has a substantial impact on the compact PWN spectrum in agreement with what was observed for this region in the  high-resolution CXO spectral maps (restricted to the  narrower, 0.5-8 keV, range; \citealt{2017JPhCS.932a2050K}). The overall (spatially-integrated) shape of the spectrum of the compact PWN is crudely consistent with the flat (in $F_\nu$) radio spectrum steepening by $\approx 0.5$ (in terms of the spectral index) in the X-ray band and having an exponential cutoff at  $\sim$50 keV. This is in agreement with the  classic picture expected for synchrotron cooling 
occurring in a stronger magnetic field within the compact PWN  (see, e.g., \citealt{2009ApJ...703..662R}). The non-uniformity  in the CXO residuals at energies $<$3 keV  could be attributed to the somewhat varying spectra ($\Gamma=1.3-1.6$) for different parts of the PWN within the $50''$ radius (as can be seen from Figure 1 in \citealt{2017JPhCS.932a2050K}). 
 The SW PWN region spectrum also does not fit a simple PL model and requires either a break near $E_b\approx3.6$ keV or exponential cutoff with $E_c\approx50$ keV.
Regardless of the model, the slope of the spectrum of the SW PWN region appears to be steeper than that of the same model fitted to the compact PWN spectrum, suggesting that the radiative losses  continue impacting particles as they travel away from the pulsar.  This behavior is often seen in PWN spectral maps (see, e.g., \citealt{2020MNRAS.498..821G}) and is expected in a simple model of radiatively cooling wind propagating away from the pulsar (see, e.g., \citealt{2006ApJ...651..237C}).   However, the spectral steepening is very gradual and the $E_c$ value does not change {\bf much}, indicating that the radiative losses are modest compared to those experienced by particles in the compact PWN region.

The maximum energy available for particle acceleration, $E_{\rm max}=e(\dot{E}/c)^{1/2}=4.6$ PeV, which corresponds to the electron Lorentz factor $\gamma_{\rm max}= 9\times10^{9}$.  Using the standard synchrotron theory    \cite{2002ASPC..271..181K}
%, 
%we can 
estimated  $B\approx 35-50~\mu$G  just downstream of the termination shock (likely associated with the double arc structure). These estimates should be applicable to the compact PWN region considered in this paper.  
For these values of magnetic field 
synchrotron 
photons  energies of $E_{\rm syn}=50$ keV  imply  electron  Lorentz factors 
$\gamma \approx 3\times10^8 (E_{\rm syn}/50\  {\rm keV})^{1/2} (B/40~{\rm \mu G})^{-1/2}$  which is only about $4\%$ of the potential drop available in the pulsar magnetosphere (for a dipolar field geometry). 
The corresponding synchrotron cooling time for these electrons is $\tau_c\approx 8 (B/40~{\rm \mu G})^{-3/2}(E_{\rm syn}/50~{\rm keV})^{-1/2}$ yrs.  The inferred maximum  Lorentz factors correspond to electron energies of $E_{e,c }\approx0.15$ PeV  is compatible  with $E_{e,c }\approx0.3^{+4}_{-0.2}$ PeV obtained  by \cite{2019A&A...627A.100H} from the joint modeling of the TeV and soft X-ray spectra from the larger region including both our compact and SW PWN regions.  Although the magnetic field used above in our estimates for the compact PWN is a factor of $3-4$ higher than that inferred by \cite{2019A&A...627A.100H}, there is no contradiction because the average magnetic field within the  much larger volume is likely to be significantly lower than that inside the compact PWN.

\section{Summary and Conclusion}

The NuSTAR data support a very complex structure of the pulse profile in 3-79 keV with at least 4 peaks per period.  
By isolating photons with phases belonging to two main peaks
 (Pk 1 and Pk2) of the pulse profile, we extracted the phase-resolved spectra and fitted them with the PL model finding photon indices of $1.10\pm 0.15$ and  $1.62\pm 0.20$, respectively. These values are substantially larger than those found earlier from RXTE data.  
 
 The fit to NuSTAR and  CXO+NuSTAR spectrum for the compact PWN requires a more complex model than a single PL and can be empirically described by a broken PL or, even better, by the exponentially cutoff PL with $E_c\approx 50$ keV.  The observed softening is likely attributed to 
 %mild 
 the synchrotron cooling in a stronger magnetic field that exists in the compact PWN.  
 The observed energies of synchrotron photons imply that at least $4\%$ 
  of the polar cap potential drop is used to accelerate pulsar wind particles  to $\approx 150$ TeV energies, which are in general agreement with those obtained from the joint soft X-ray and TeV spectral modeling by \cite{2019A&A...627A.100H}.

\medskip\noindent{\bf Acknowledgments:}
We thank Matthew Kerr for providing the Fermi Vela ephemeris. We also thank Matthew Kerr, Paul Ray, and Craig Markwardt for useful conversations regarding pulsar timing. OK is  grateful to Sasha Philippov for discussions about pulsar emission models.  Finally, we would like to thank the anonymous referee for the careful reading of our paper and insightful comments that helped to improve its quality. 
Support for this work was provided by the National Aeronautics and Space Administration through
the NuSTAR award 80NSSC21K0024. J. H. acknowledges support from NASA under award number 80GSFC21M0002.

\end{document}